\documentclass{iaus238}
\usepackage{graphicx}

\newcommand{\myemph}[1]{\emph{#1},}
\title[Black holes and magnetic fields] %% give here short title %%
{Black holes and magnetic fields}

\author[J.~Bi\v c\'ak, V. Karas \& T.~Ledvinka] %% give here short author list %%
{Ji\v r\'\i \ Bi\v c\'ak,$^1$ Vladim\'\i r Karas$^2$ \and Tom\'a\v s\ Ledvinka$^1$}

\affiliation{
$^1$Institute of Theoretical Physics, Faculty of Mathematics and Physics, Charles University, V~Hole\v{s}ovi\v{c}k\'ach~2, CZ-18000~Prague, Czech Republic\\[\affilskip]
$^2$Astronomical Institute, Academy of Sciences, Bo\v{c}n\'{\i}~II, CZ-14131~Prague, Czech Republic\\[\affilskip]
email: bicak@mbox.troja.mff.cuni.cz, vladimir.karas@cuni.cz, ledvinka@mbox.troja.mff.cuni.cz}

\pubyear{2007}
\volume{238}  %% insert here IAU Symposium No.
\date{}
\jname{Black Holes: from Stars to Galaxies -- across the Range of Masses}
\editors{V. Karas \& G. Matt, eds.}
\begin{document}

\maketitle

\begin{abstract}
Stationary axisymmetric magnetic fields are expelled from outer horizons
of black holes as they become extremal. Extreme black holes exhibit
Meissner effect also within exact Einstein--Maxwell theory and in string
theories in higher dimensions. Since maximally rotating black holes are
expected to be astrophysically most important, the expulsion of the
magnetic flux from their horizons represents a potential threat to an
electromagnetic mechanism launching the jets at the account of
black-hole rotation.
\end{abstract}

\firstsection % if your document starts with a section,
\section{Introduction}
The exact mechanism of formation of highly relativistic jets from
galactic nuclei and microquasars remains unknown. Four ways by which a
black hole or its accretion disk could power two opposite jets  are
indicated in figure~1 (taken from the \emph{Czech} edition of Kip
Thorne's popular book to indicate how, as compared with the last 1967
IAU meeting in Prague, black holes domesticated even in central Europe):
(a)~wind from the disk may blow a bubble in a spinning gas cloud and
hot gas makes the orifices through which jets are shot out; (b)~the
surface of the puffed rotating disk forms funnels which collimate the
wind; (c)~magnetic field lines anchored in the disk are spinning due to
disk's rotation and push plasma to form jets; (d)~magnetic lines
threading through the hole are forced to spin by the ``rotating
geometry'' and push plasma outwards along the rotation axis.

\begin{figure}[ht]
%\begin{center}
\includegraphics[width=0.78\textwidth]{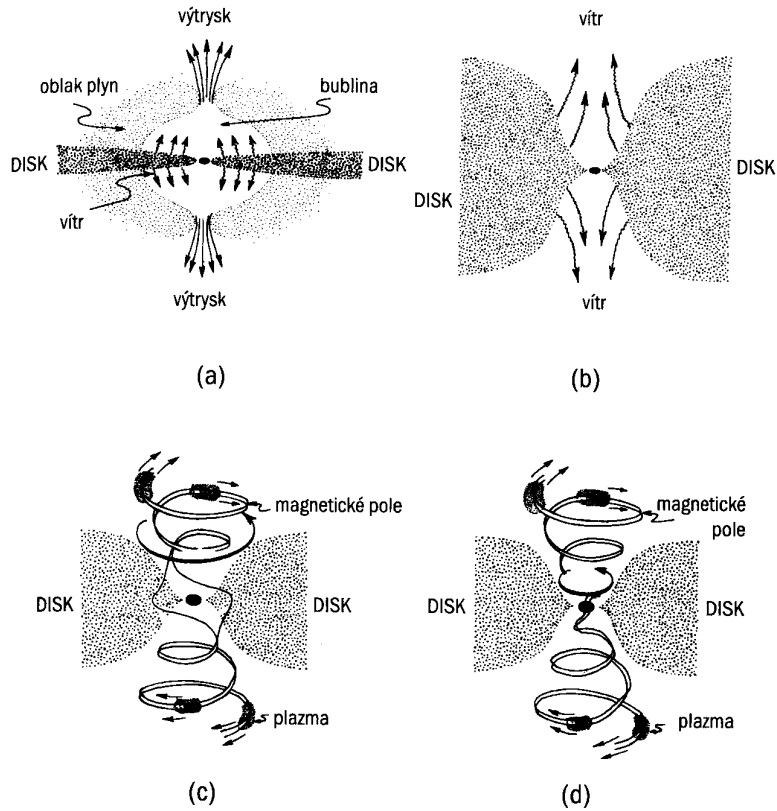}
%\includegraphics[width=0.49\textwidth,bb=21 187 324 340,clip=true]{abcdbw.eps}
%\includegraphics[width=0.49\textwidth,bb=21 21 324 164,clip=true]{abcdbw.eps}
%\end{center}
\hfill
\parbox[b]{0.21\textwidth}{\caption{Four mechanisms of jet formation 
(drawing from ``\v{C}ern\'e d\'{\i}ry a  zborcen\'y \v{c}as'', the Czech 
edition of ``Black Holes and Time Warps'', by K.~S. 
Thorne 2004).\newline\newline}\label{fig:fromthorne}}
\end{figure}

The last way, the Blandford--Znajek mechanism, is considered to be the
most relevant. The field brought into the innermost region and onto the
black hole from the outside has clean field structure while the field
in/around the disk is expected to be quite chaotic (Thorne {\etal} 1986).
The estimated power radiated out from the ``load'' regions farther away
from the hole is
\begin{equation}
\Delta L_{\rm max} \simeq 
\left[ 10^{45} {\rm erg\over sec}\right] 
\left[a\over M \right]^2 
\left[M\over 10^{9}M_\odot \right]^2 
\left[B_n\over 10^4 G \right]^2.
\end{equation}
Here $a\,\equiv\,J/M$ is the hole's angular momentum per unit mass 
(velocity of light $c=1$, gravitational constant $G=1$), $B_n$ is the
normal magnetic field at the horizon. Hence, the highest power is
achieved when the hole is ``extreme'', i.e., rotating with maximal
angular momentum or approaching the $a=M$ limit, and $B_n$ determining
the magnetic flux across the horizon as high as possible. 

\begin{figure}[ht]
 \begin{center}
 \includegraphics[width=0.35\textwidth,bb=142 520 342 720,clip=true]{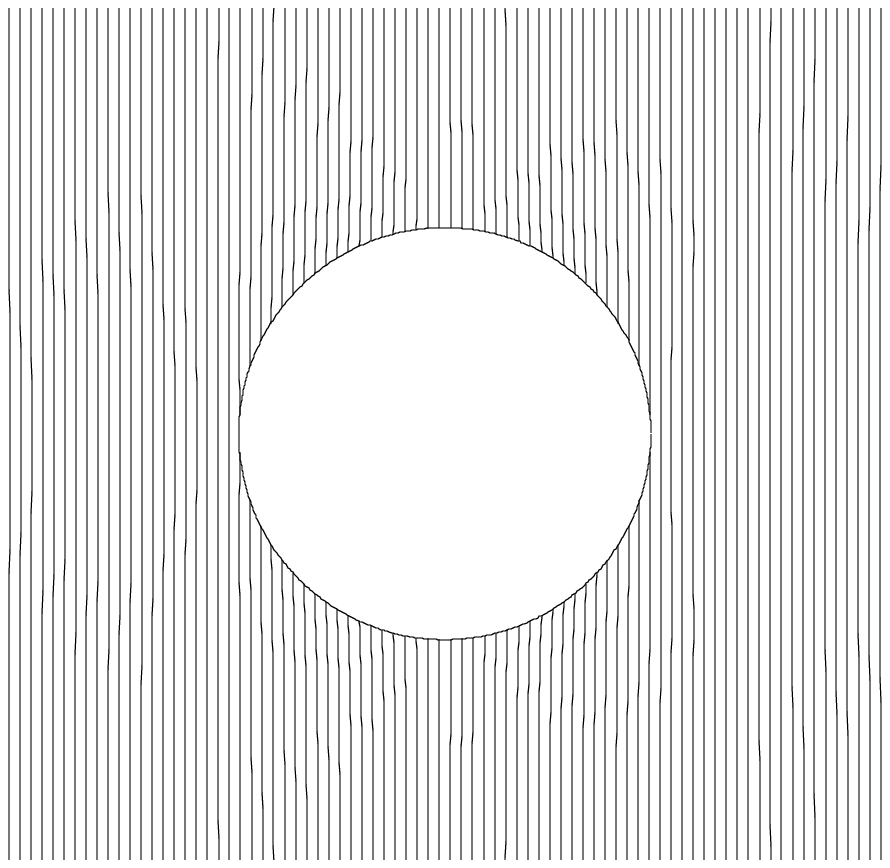}
~~~~~~~~~~~~~~
 \includegraphics[width=0.35\textwidth,bb=142 520 342 720,clip=true]{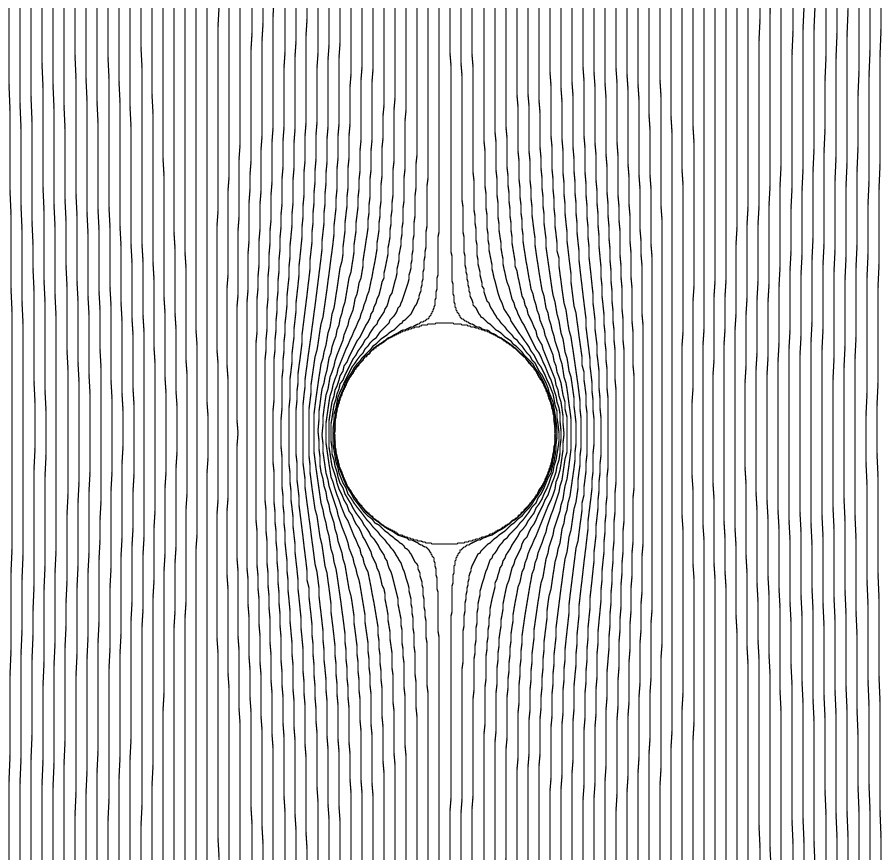}
 \end{center}
  \caption{Field lines of the test magnetic field uniform at infinity
and aligned with hole's rotation axis. Two cases with $a=0.5M$ (left)
and $a=M$ (right) are shown.}\label{fig:mgfields}
%\end{figure}
~\\
%\begin{figure}[h]
 \begin{center}
 \includegraphics[width=0.35\textwidth,bb=142 230 490 577,clip=true]{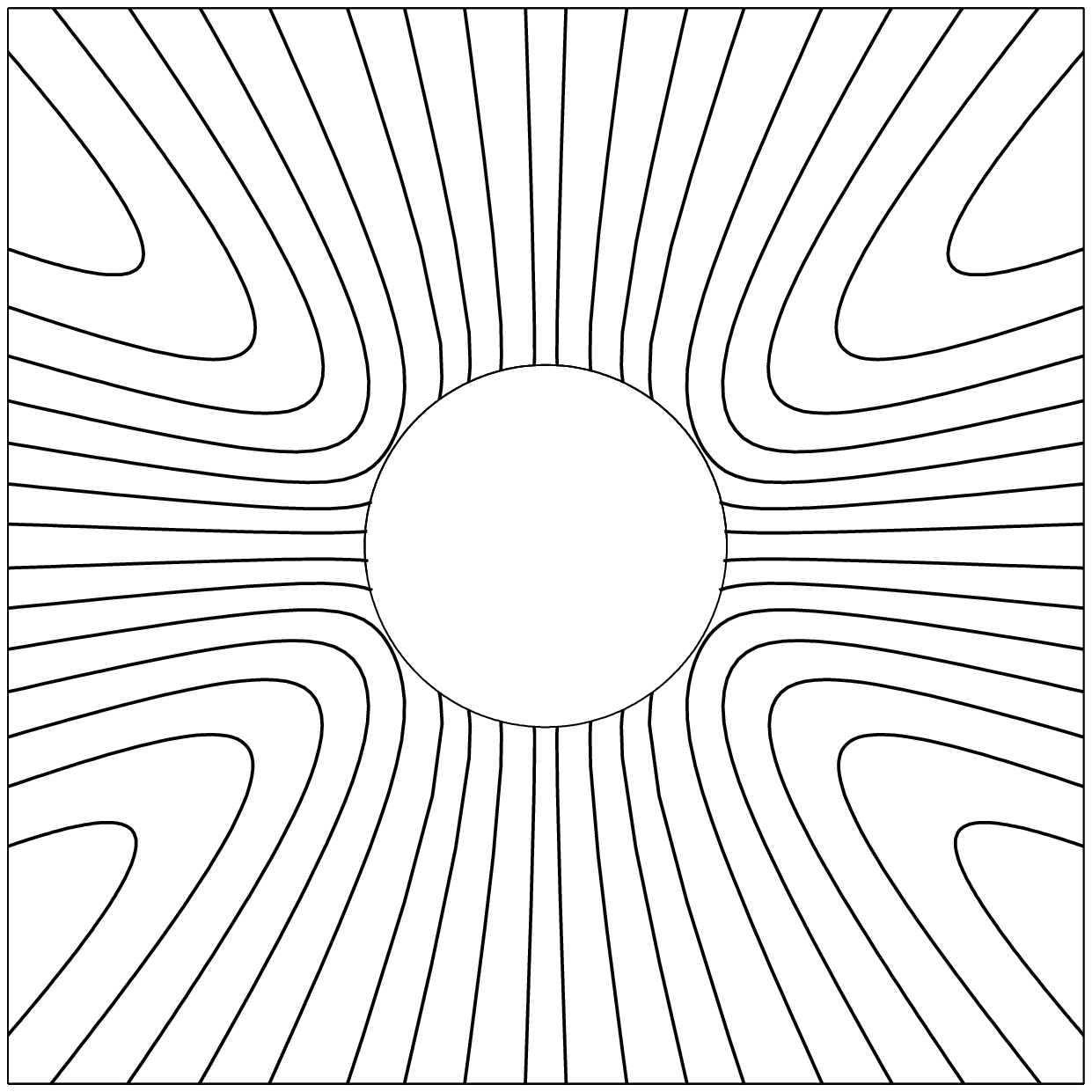}
~~~~~~~~~~~~~~
 \includegraphics[width=0.35\textwidth,bb=142 230 490 577,clip=true]{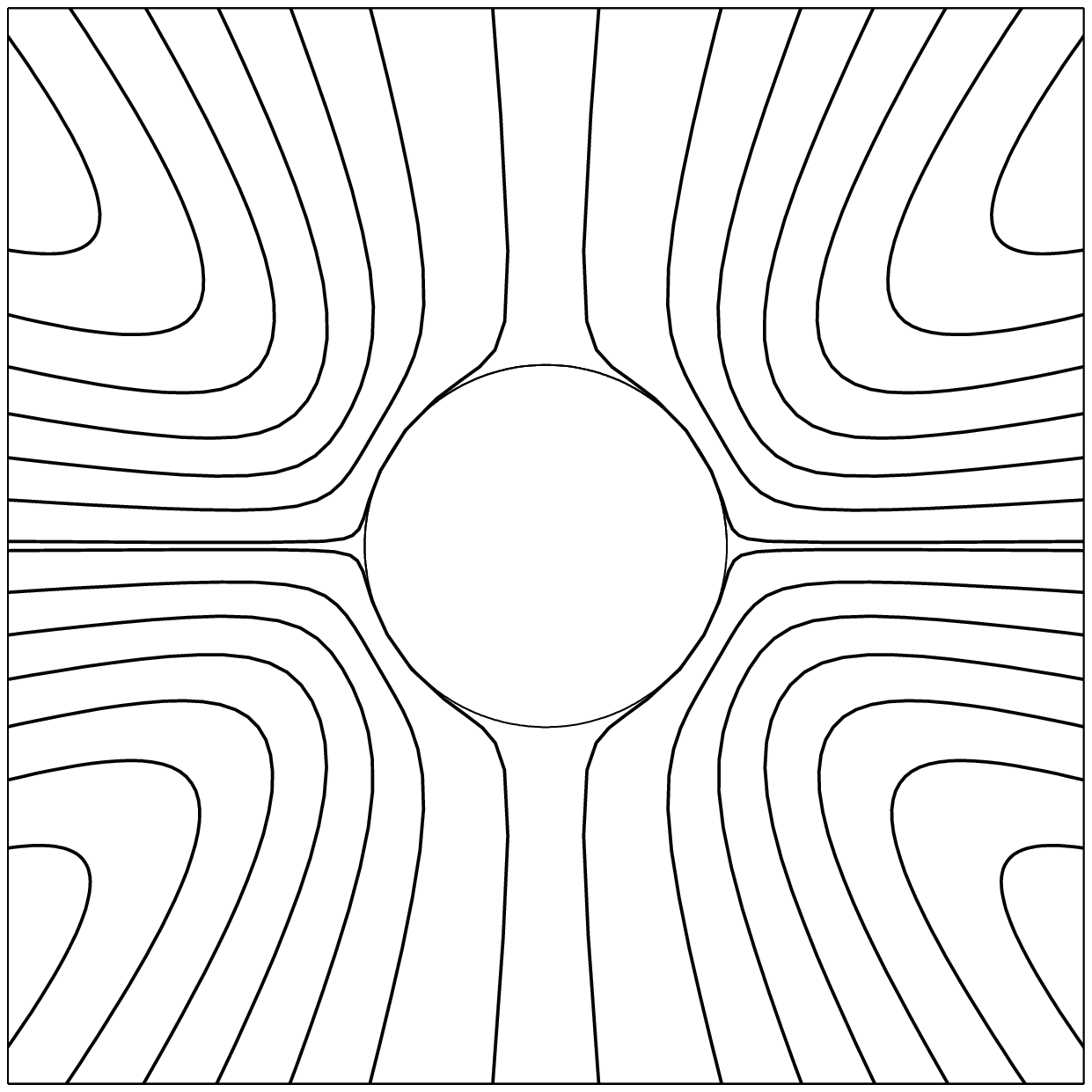}
 \end{center}
  \caption{Field lines of the electric field induced by the ``rotating
geometry" of Kerr black hole in asymptotically uniform test magnetic
field. $a=0.95 M$ (left), and $a=M$ (right). }\label{fig:elfields}
\end{figure}

\section{Meissner effect}
The main purpose of this contribution is to point out that these two
aspects go one against the other: black holes approaching extremal
states exhibit a ``Meissner effect'' -- they expel external vacuum
stationary (electro)magnetic fields.

To see this effect in the simplest situation, consider the magnetic 
test field $B_0$ which is uniform at infinity and aligned with hole's
rotation axis. Solution of Maxwell's equations on the background
geometry of a rotating (Kerr) black hole with boundary condition of
uniformity at infinity and finiteness at the horizon yields the field
components; from these the lines of force are defined as lines tangent
to the Lorentz force experienced by test magnetic/electric charges at
rest with respect to locally non-rotating frames (preferred by the Kerr
background field). The field lines are plotted in figure~2 for $a=0.5M$
and in extreme case $a=M$. Notice that only weak expulsion occurs in the
former case. There is a simple analytic formula for the flux across the
hemisphere of the horizon: $\Phi=B_0 \pi r_+^2 (1-a^4/r_+^4)$, where
$r_+=M+(M^2-a^2)^{1/2}$
(King \etal\ 1975; Bi\v{c}\'ak \& Jani\v{s} 1985). 

As a consequence of the coupling of magnetic field to frame-dragging
effects of the Kerr geometry the electric field of a quadrupolar nature
arises. Its field lines are shown in figure~3. Again the flux expulsion
takes place. While even with $a=0.95M$ it is still not very distinct, 
the expulsion becomes complete in the extreme case. 

\begin{figure}[ht]
 \begin{center}
 \includegraphics[width=0.35\textwidth]{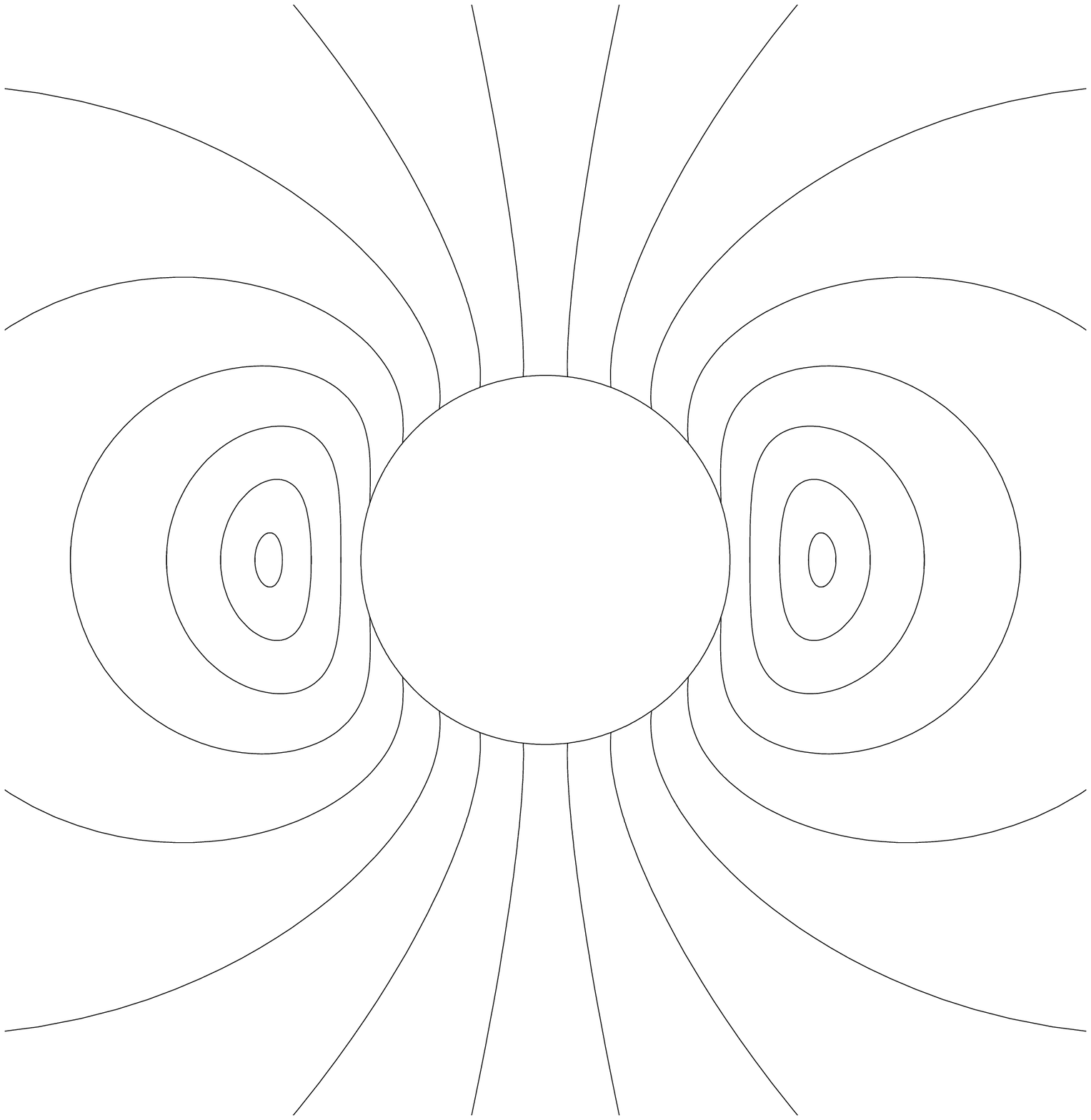}
~~~~~~~~~~~~
 \includegraphics[width=0.35\textwidth]{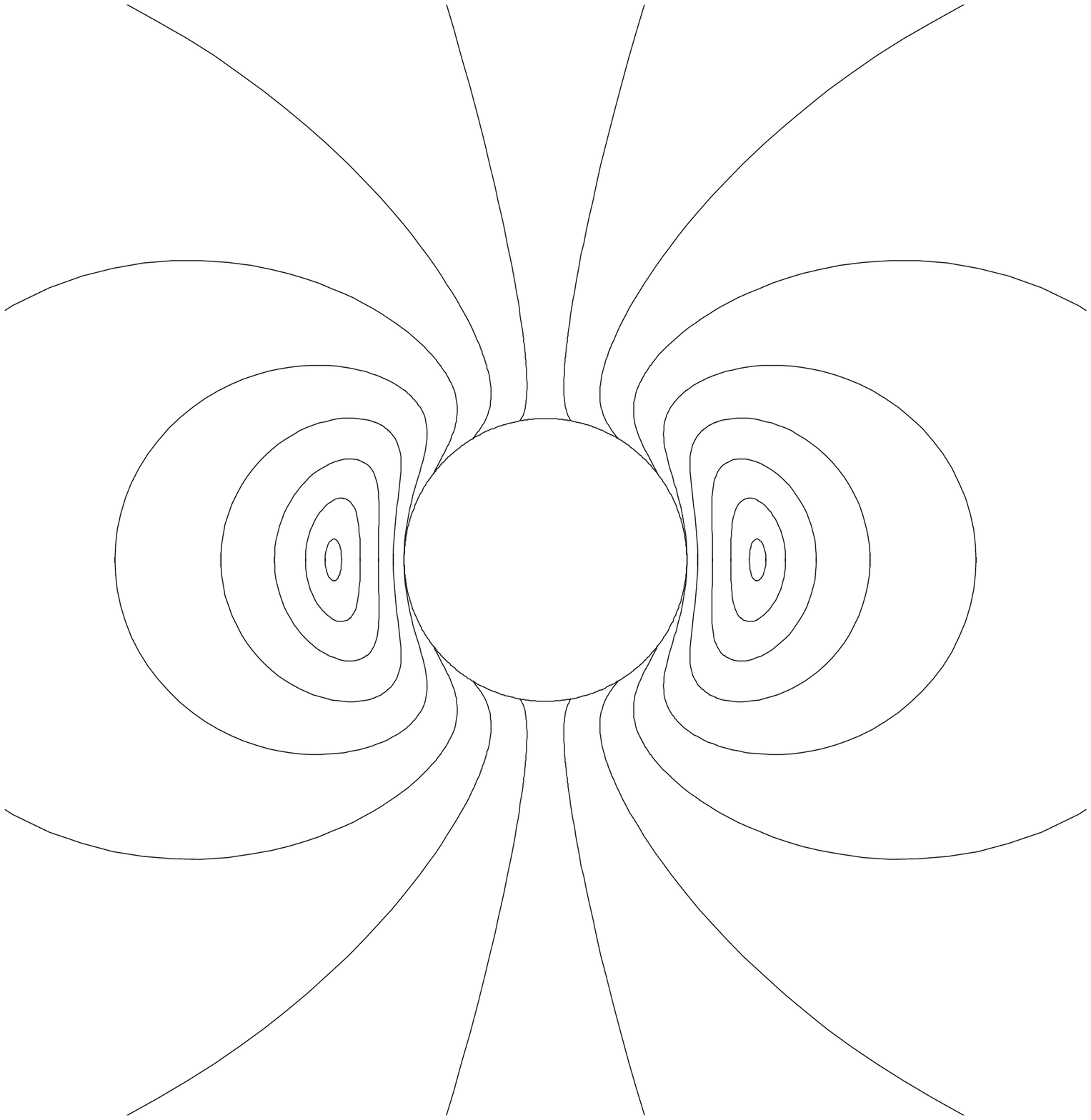}
 \end{center}
  \caption{Field lines of the magnetic field of a current loop in the
equatorial plane of Kerr black hole located at $r=1.5~r_+$. Two cases
with $a=0.9 M$ (left) and $a=0.995 M$ (right) are shown.
}\label{fig:currentloop} %\end{figure} ~\\ %\begin{figure}[h]
 \begin{center}
 \includegraphics[width=0.35\textwidth]{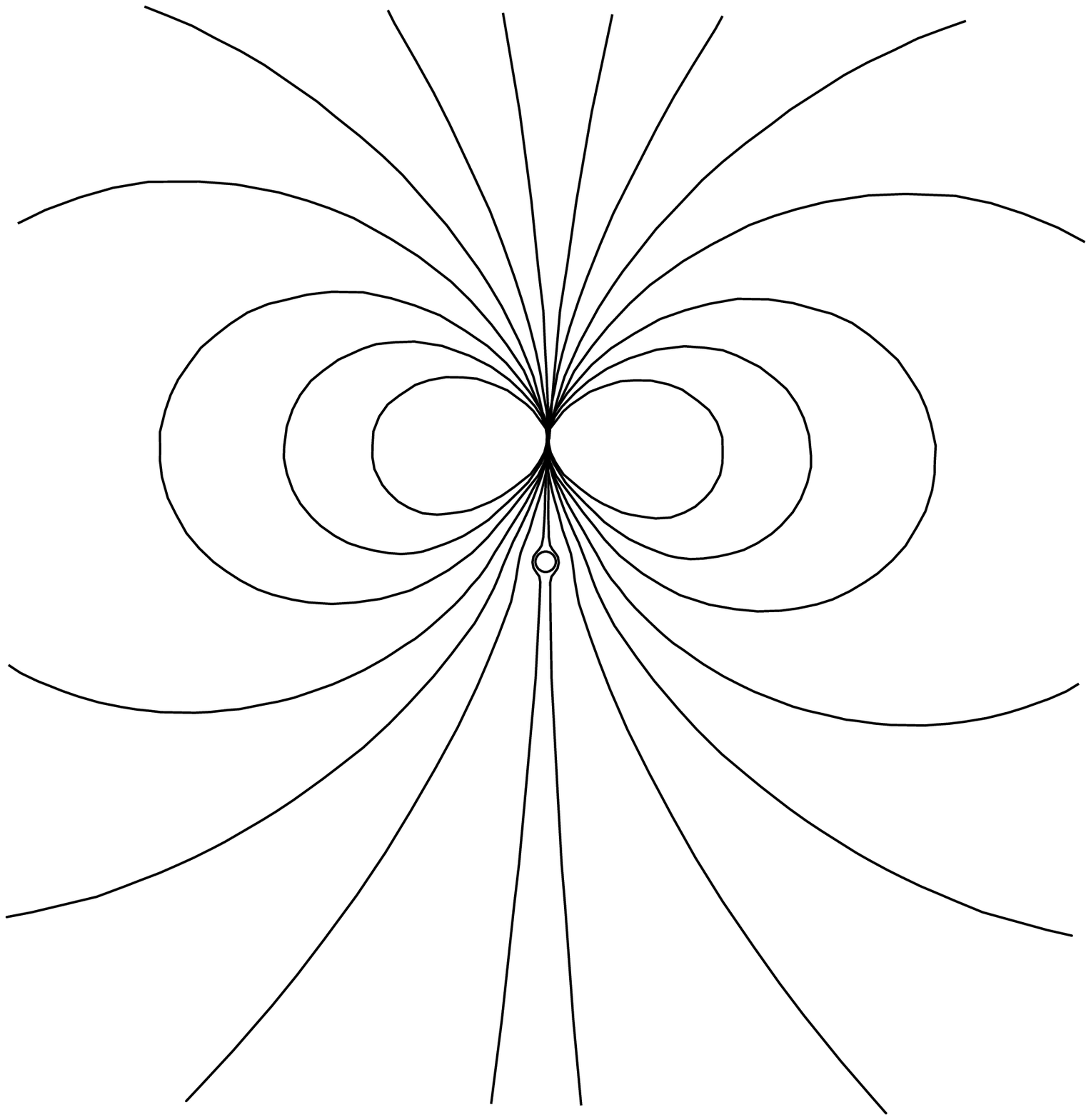}
~~~~~~~~~~~~
 \includegraphics[width=0.35\textwidth]{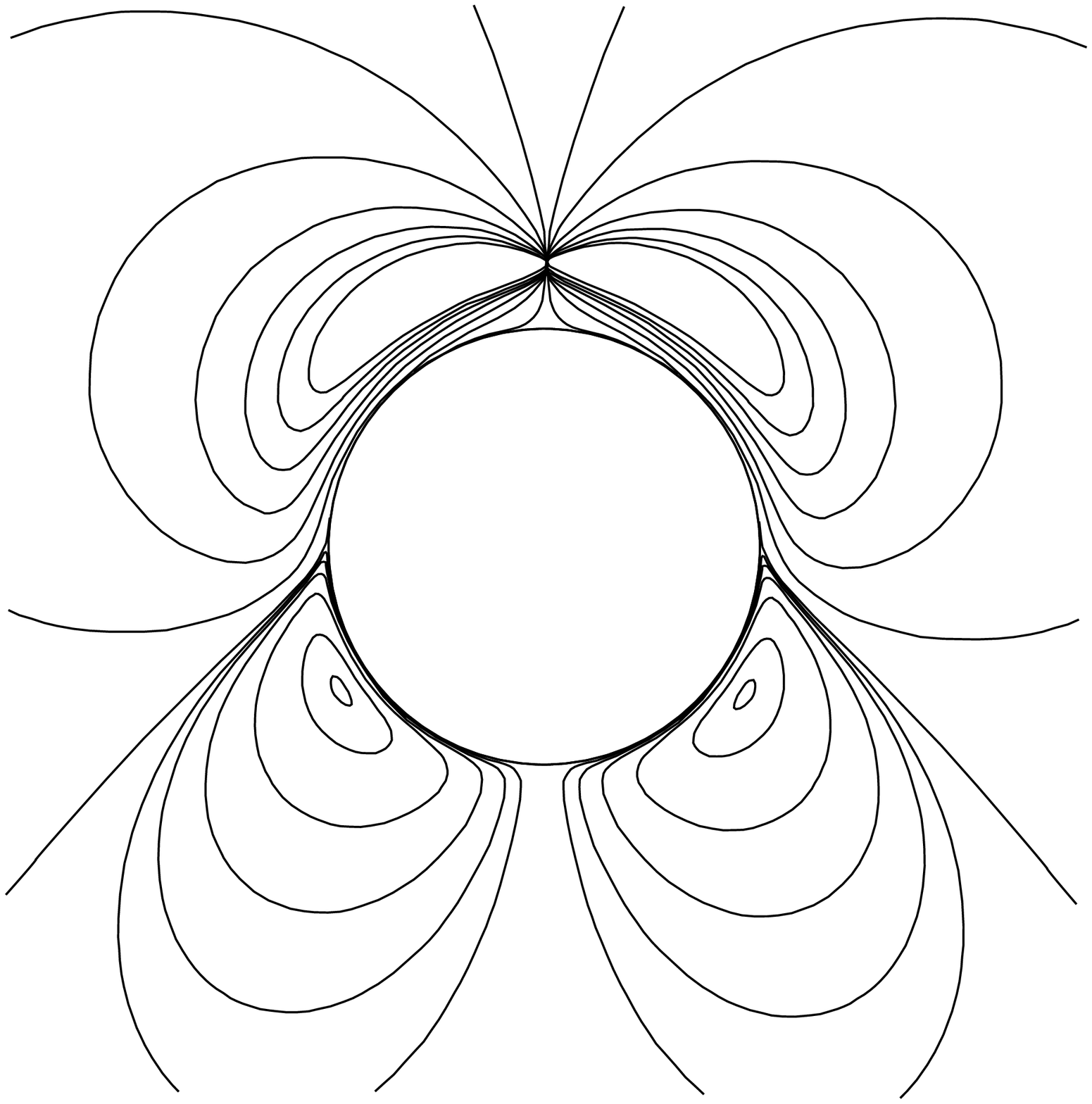}
 \end{center}
  \caption{Field lines of the magnetic dipole placed far away from
the extreme Reissner--Nordstr\"{o}m black hole ($e=M$, left panel), 
and close to the hole (right panel). }\label{fig:butterfly}
\end{figure}

One can demonstrate that total flux expansion takes place for all
axisymmetric stationary fields around  a rotating black hole
(Bi\v{c}\'ak \& Jani\v{s} 1985; Bi\v{c}\'ak \& Ledvinka 2000). In
figure~4 the field lines of a current loop in the equatorial plane
are shown. 

The Meissner-type effect arises also for charged (Reissner--Nordstr\"om)
black holes. Although extremely charged  black holes ($e^2=M^2$) are
probably not important astrophysically  they may be significant in
fundamental physics (as very special supersymmetric BPS states mass of
which does not get any quantum corrections). Since electromagnetic
perturbations are in general coupled to gravitational perturbations, the
resulting formalism is involved. Nevertheless, one may construct
explicit solutions, at least in stationary cases. From these the
magnetic field lines follow as in the Kerr case. The magnetic field
lines  of a dipole far away from the hole look like in a flat space
(Fig.~5a), however, when the dipole is close to the horizon, the
expulsion in the extreme case is evident (Fig.~5b). Due to the coupling
of perturbations closed field lines appear without any electric current
inside; see \cite{BiDvo} for details. 

There exist exact models (exact solutions of the Einstein--Maxwell
equations) representing in general rotating, charged black holes
immersed in an axisymmetric magnetic field. The expulsion takes place
also within this exact framework -- see \cite{BiKa}, \cite{KaVo},
\cite{KaBu}.

Very recently the Meissner effect was demonstrated for extremal
black-hole solutions in higher dimensions in string theory and
Kaluza--Klein theory. The question of the flux expulsion from the
horizons of extreme black holes in more general frameworks is not yet
understood properly. The authors of \cite{CEG} ``believe this to be a
generic phenomenon for black holes in theories with more complicated
field content, although a precise specification of the dynamical
situations where this effect is present seems to be out of reach.''

The flux expulsion does \textit{not} occur for the fields which are not
axisymmetric. In figure~6 the field lines are constructed for fields
asymptotically uniform and perpendicular to the axis of an extreme Kerr
black hole. There is an angle $\delta_{\rm max}\sim-63^\circ$ for which
the flux of $B_1$ component across a hemisphere is maximal, $ \Phi_{\rm
max}\sim 2.25B_1\pi r^2_+$. This is the effect of the rotating geometry
(for detailed description,  see Bi\v{c}\'ak \& Jani\v{s} 1985;
Dov\v{c}iak {\etal} 2000). In these ``misaligned'' situations, there is an
angular momentum flux to infinity, so such conditions are not
stationary.

Naturally, the shape of magnetic and electric field lines depends on the
choice of observers, however, the whole discussion can be cast
in equivalent and invariant form by employing surfaces of constant flux
in which field lines reside. In this way, figure~7 demonstrates that 
the cross-section of the black hole for the capture of non-aligned
magnetic fields indeed does not vanish as the hole rotation aproaches
the extreme value.

\begin{figure}[ht]
 \begin{center}
 \includegraphics[height=.33\textwidth]{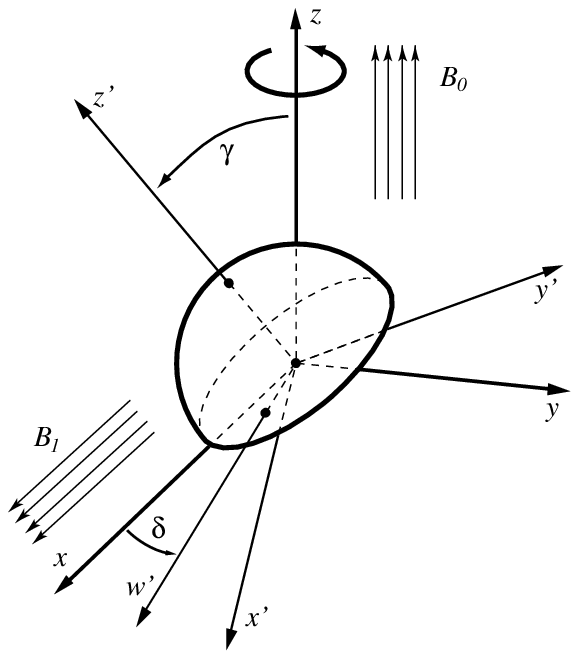}~~~\hfil
\includegraphics[height=.33\textwidth]{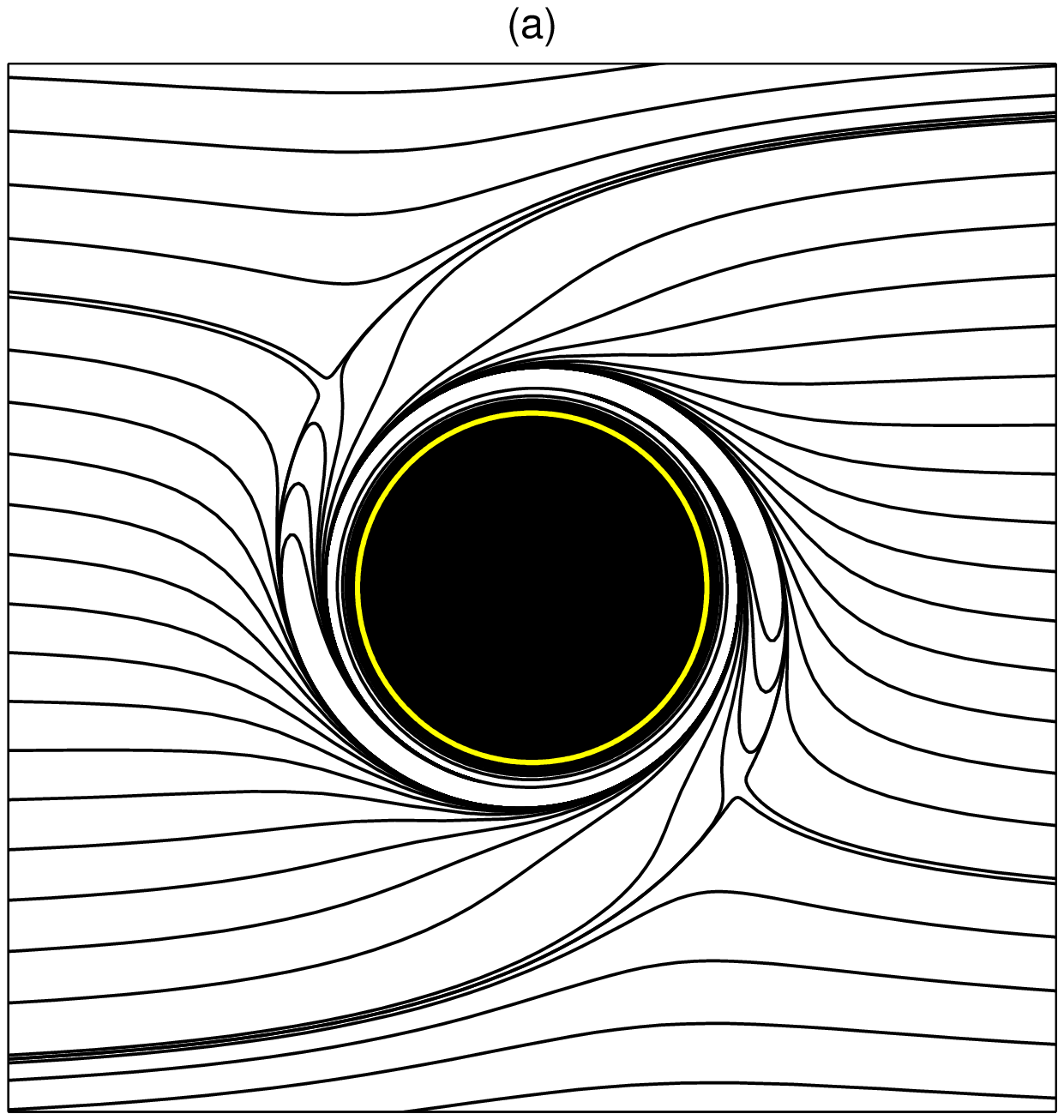}~~~\hfil
\includegraphics[height=.33\textwidth]{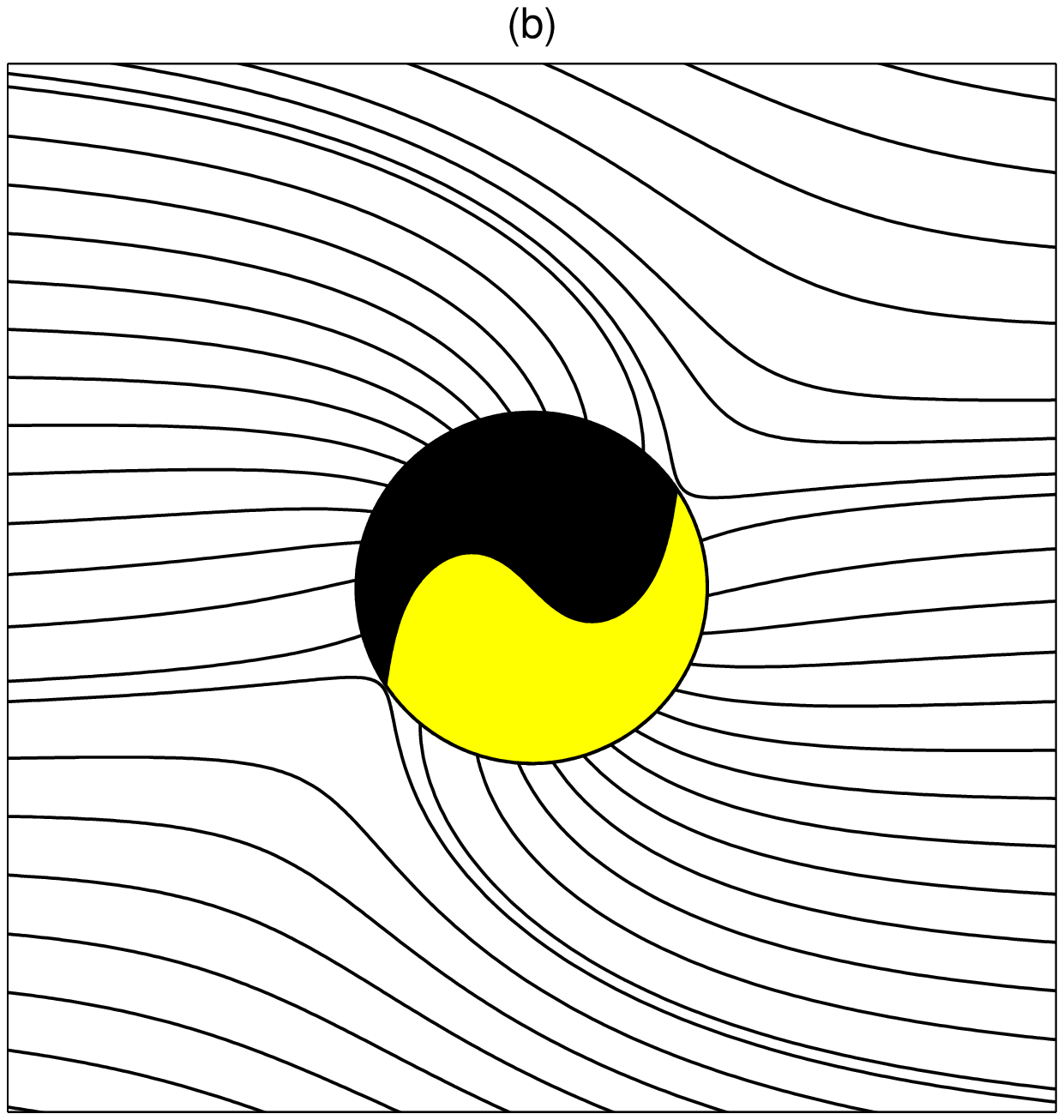}
 \end{center}
  \caption{Lines of the magnetic field which is asymptotically uniform 
and perpendicular to the rotation axis. The equatorial plane is shown as
viewed from top, i.e.\ along the rotation axis, (a)~in the frame of zero
angular momentum observers orbiting at constant radius; (b)~in the frame
of  freely falling observers. In the panel (b), two regions of 
ingoing/outgoing lines are distinguished by different levels of shading 
of the horizon (the hole rotates counter-clockwise,
$a=M$).}\label{fig:nonaligned}
\end{figure}

\begin{figure}[th]
\begin{center}
\includegraphics[width=0.5\textwidth]{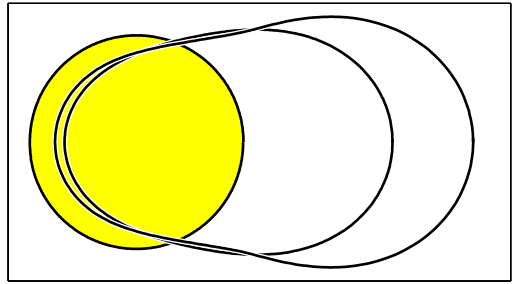}
\end{center}
\caption{The cross-sectional area for the capture of magnetic field lines 
(asymptotically uniform magnetic field perpendicular to the rotation axis).
The three curves correspond to different values of the black-hole 
angular momentum: $a=0$ (the black hole cross-section is a perfect circle
and its projection coincides with the black-hole horizon of radius $2M$, 
indicated here by shading), $a=0.95M$, and $a=M$ (the most deformed 
shape refers to the maximally rotating case). Non-vanishing cross-sectional
area for the extreme case demonstrates that the Meissner-type expulsion of the 
magnetic field does not operate on non-axisymmetric fields. 
The hole's rotation axis is vertical and the magnetic field lines are 
pointing ``towards us'' (see 
Dov\v{c}iak \etal\ 2000 for details).}\label{fig:nonaligned2}
\end{figure}

\section{Recent progress, open problems}

Can these important properties of electromagnetic fields in neighborhood
of black holes be relevant in astrophysical conditions? Recently
remarkable progress in axisymmetric simulations of rotating black holes 
surrounded by magnetized plasma based on general-relativistic MHD has
been achieved by various authors, in particular by C.~Gammie,
S.~Komissarov, J.~McKinney, J.~Krolik, D.~Uzdensky, H.~Kim, A.~Aliev and
others (see the E-print archive for references). Such simulations could
bring an answer although some new analytic insight may also be required.
For example, \cite{McKinney} remark ``we see no sign of expulsion of
flux from the horizon... It is possible that we have not gone close
enough to $a/M=1$.'' They have $a/M=0.938$ and, indeed, in view of
simplest (though just vacuum) situations illustrated in figures 2 and 3
this value may still be far from $1$ to see the effect.

An interesting issue arises in connection with  the ``black-hole
membrane paradigm'' of Thorne {\etal} (1986).  For an extreme black hole
the proper distance from any $r>r_+$ to the horizon is logarithmically
infinite, so one might tend to explain the vanishing of magnetic flux by
this fact. However, although the flux across a ``stretched horizon'' (at
a finite distance) is non-vanishing even in the extreme case, it depends
on where the stretched horizon is located. It turns out, (see section 4
of \cite{BiLe}) that for any $\epsilon>0$ one can find such a stretched
horizon that the flux is less then $\epsilon$. This suggests the
following question: does the power in the Blandford--Znajek model arise
from regions with ``relatively large $\sqrt{-g_{00}}$'' in near extreme
cases?

Another potential obstacle for Blandford--Znajek mechanism to operate
efficiently is a low value of magnetic field brought onto the black hole
from an accretion disk in realistic situations. Most recently \cite{RGB}
obtained an encouraging result on trapping of magnetic flux by the
plunge region of a black hole accretion disk. Their analysis is so far 
limited to slowly rotating black holes; it does not use general
relativity; and it depends crucially on the chosen boundary conditions.
It is not clear of how large is the disk region from which flux can be
dragged inwards.

The main open issue can be stated simply: ``Do extremely rotating black
holes produce relativistic jets?'' A compelling answer may be out of
reach for some time yet.

\begin{acknowledgments}
We acknowledge the continued support from the Czech Science Foundation 
(ref.\ 202/06/0041), as well as the Centre for Theoretical Astrophysics
and the Research Program of the Czech Ministery of Education.
\end{acknowledgments}

%\discuss{Andrew King}{Comment: A misaligned magnetic field tries to reach 
%axisymmetric configuration. It is interesting to notice that the
%timescale for that is comparable to the Blandford-Znajek timescale.}

%\discuss{Ji\v{r}\'{\i} Bi\v{c}\'ak}{Yes indeed, the tendency to align the magnetic 
%field with the black hole rotation axis is natural here, since in a misaligned 
%case there is the angular momentum flux out of the system. Of course, even 
%in case of the misaligned geometry the magnetic field component parallel to the 
%axis becomes expelled out of the extreme Kerr black hole.}
%\cleardoublepage 
\end{document}